\documentstyle{mn}

\newif\ifAMStwofonts \AMStwofontstrue

\ifoldfss
  \ifCUPmtlplainloaded \else
    \NewTextAlphabet{textbfit} {cmbxti10} {}
    \NewTextAlphabet{textbfss} {cmssbx10} {}
    \NewMathAlphabet{mathbfit} {cmbxti10} {} % for math mode
    \NewMathAlphabet{mathbfss} {cmssbx10} {} %  "   "    "
  \fi
  \ifAMStwofonts
    \ifCUPmtlplainloaded \else
      \NewSymbolFont{upmath} {eurm10}
      \NewSymbolFont{AMSa} {msam10}
      \NewMathSymbol{\upi}     {0}{upmath}{19}
      \NewMathSymbol{\umu}     {0}{upmath}{16}
      \NewMathSymbol{\upartial}{0}{upmath}{40}
      \NewMathSymbol{\leqslant}{3}{AMSa}{36}
      \NewMathSymbol{\geqslant}{3}{AMSa}{3E}

      \let\leq=\leqslant 
      \let\geq=\geqslant 
    \fi
  \fi
\fi % End of OFSS

\ifnfssone
  \newmathalphabet{\mathit}
  \addtoversion{normal}{\mathit}{cmr}{m}{it}
  \addtoversion{bold}{\mathit}{cmr}{bx}{it}
  \newmathalphabet{\mathbfit} % math mode version of \textbfit{..}
  \addtoversion{normal}{\mathbfit}{cmr}{bx}{it}
  \addtoversion{bold}{\mathbfit}{cmr}{bx}{it}
  \newmathalphabet{\mathbfss} % math mode version of \textbfss{..}
  \addtoversion{normal}{\mathbfss}{cmss}{bx}{n}
  \addtoversion{bold}{\mathbfss}{cmss}{bx}{n}
  \ifAMStwofonts
    \ifCUPmtlplainloaded \else
      \UseAMStwoboldmath
      \makeatletter
      \new@mathgroup\upmath@group
      \define@mathgroup\mv@normal\upmath@group{eur}{m}{n}
      \define@mathgroup\mv@bold\upmath@group{eur}{b}{n}
      \edef\UPM{\hexnumber\upmath@group}
      \new@mathgroup\amsa@group
      \define@mathgroup\mv@normal\amsa@group{msa}{m}{n}
      \define@mathgroup\mv@bold\amsa@group{msa}{m}{n}
      \edef\AMSa{\hexnumber\amsa@group}
      \makeatother
      \mathchardef\upi="0\UPM19
      \mathchardef\umu="0\UPM16
      \mathchardef\upartial="0\UPM40
      \mathchardef\leqslant="3\AMSa36
      \mathchardef\geqslant="3\AMSa3E

      \let\leq=\leqslant 
      \let\geq=\geqslant 
    \fi
  \fi
\fi % End of NFSS release 1

\ifnfsstwo
  \DeclareMathAlphabet{\mathbfit}{OT1}{cmr}{bx}{it}
  \SetMathAlphabet\mathbfit{bold}{OT1}{cmr}{bx}{it}
  \DeclareMathAlphabet{\mathbfss}{OT1}{cmss}{bx}{n}
  \SetMathAlphabet\mathbfss{bold}{OT1}{cmss}{bx}{n}
  \ifAMStwofonts
    \ifCUPmtlplainloaded \else
      \DeclareSymbolFont{UPM}{U}{eur}{m}{n}
      \SetSymbolFont{UPM}{bold}{U}{eur}{b}{n}
      \DeclareSymbolFont{AMSa}{U}{msa}{m}{n}
      \DeclareMathSymbol{\upi}{0}{UPM}{"19}
      \DeclareMathSymbol{\umu}{0}{UPM}{"16}
      \DeclareMathSymbol{\upartial}{0}{UPM}{"40}
      \DeclareMathSymbol{\leqslant}{3}{AMSa}{"36}
      \DeclareMathSymbol{\geqslant}{3}{AMSa}{"3E}

      \let\leq=\leqslant 
      \let\geq=\geqslant 
    \fi
  \fi
\fi % End of NFSS release 2

\ifCUPmtlplainloaded \else
  \ifAMStwofonts \else % If no AMS fonts
    \def\upi{\pi}
    \def\umu{\mu}
    \def\upartial{\partial}
  \fi
\fi

\title{The symbiotic star CH Cygni. I. Non-thermal bipolar jets}

\author[M. M. Crocker et al.]  {M. M. Crocker$^1$, R. J. Davis$^1$,
       S. P. S. Eyres$^2$, M. F. Bode$^3$,  \newauthor
       A. R. Taylor$^4$, A. Skopal$^5$ and H. T. Kenny$^6$ \\
       $^1$Jodrell Bank Observatory, University of Manchester, Macclesfield, Cheshire, SK11 9DL, UK.\\
       $^2$Centre for Astrophysics, University of Central Lancashire, Preston, PR1 2HE, UK\\
       $^3$Liverpool John Moores University, Twelve Quays House, Egerton Wharf, Birkenhead, CH41 1LD, UK\\ 
       $^4$The Department of Physics and Astronomy, The University of Calgary, 2500 University Dr. N.W., Calgary, Alberta, T2N 1N4, Canada\\ 
       $^5$Astronomical Institute, Slovak Academy of Sciences, 059~60 Tatransk\'{a} Lomnica, Slovakia\\ 
       $^6$Department of Physics, Royal Military College of Canada, P.O. Box 17000, Stn Forces, Kingston, ON, K7K 7B4, Canada
}

\date{Accepted ????.  Received ????; in original form ????}

\pagerange{\pageref{firstpage}--\pageref{lastpage}} \pubyear{2001}

\begin{document}

\maketitle

\label{firstpage}

\begin{abstract}
VLA surface brightness and spectral index maps of the evolving
extended emission of the triple symbiotic star CH Cygni are
presented. These are derived from observations at 4.8, 8.4 and  14~GHz
between 1985 and 1999. The maps are dominated by thermal emission
around the central bright peak of the nebula, but we also find
unambiguous non--thermal emission associated with the extended
regions.  Our observations confirm that this is a jet. The central
region has been associated with the stellar components through HST
imaging. If the jets are due to ejection events at outburst, expansion
velocities are  consistent with those from other measurement
methods. We propose that the non--thermal emission is caused by
material ejected in the bipolar jets interacting with the
circumstellar wind envelope. The resulting shocks lead to local
enhancements in the magnetic  field from the compact component of the
order 3~mG.
\end{abstract}


\section{Introduction}

Symbiotic stars occupy an extreme and relatively poorly understood
region of the binary classification scheme. The name was coined by
Merril (1941) to describe stars which appeared to  have a combination
spectra: that of high excitation lines usually associated with a hot,
ionized nebula superimposed on a cool continuum with prominent
absorption features consistent with a late--type star. At present they
are understood to be interacting binaries (with orbital periods of a
few to tens of years) consisting of a cool giant losing material
mostly via the stellar wind  and a hot, luminous compact object which
ionises a portion of the cool component wind (Seaquist et al., 1984,
hereafter the STB model). Such a state  of affairs represents the
so--called {\em quiescent phase}, which can be interrupted by periods
of  activity. The {\em active phases} start with an eruption of the
hot star, an event indicated photometrically by an increase of the
star's brightness by 2-6\,mag, and/or spectroscopically by high
velocity broad emission features from the central star. Both radio and
HST imaging can directly resolve the remnants of such dramatic events
(Hack \& Parsece, 1993; Eyres et al. 1995; Kenny et al. 1996; Richards 
et al., 1999; Eyres et al., 2001; Eyres et al. in preparation, hereafter 
Paper II).

The symbiotic star CH\,Cygni displays particularly complex
behaviour. Optical spectroscopic studies by Miko{\l}ajewski et al
(1987) showed that the orbits of the stars within the system are
likely to be coplanar and eclipsing, with eclipses separated by around
5700 days.  Later studies (Miko{\l}ajewski et al, 1990 and references
therein) confirmed this period.

Further spectral studies of the system (Hinkle et al. 1993) suggested that
CH\,Cyg is probably a triple-star system consisting of an inner
756-day period binary which is orbited by an
unseen G-K dwarf on a 5300~day orbit. Further photometric studies
(Skopal et al., 1996) discovered short, 756 day, period eclipses which
show that all three stars in the system are likely to be in coplanar
orbits.

\begin{figure*}
\begin{picture}(200,220)
 \put(0,0){\includegraphics{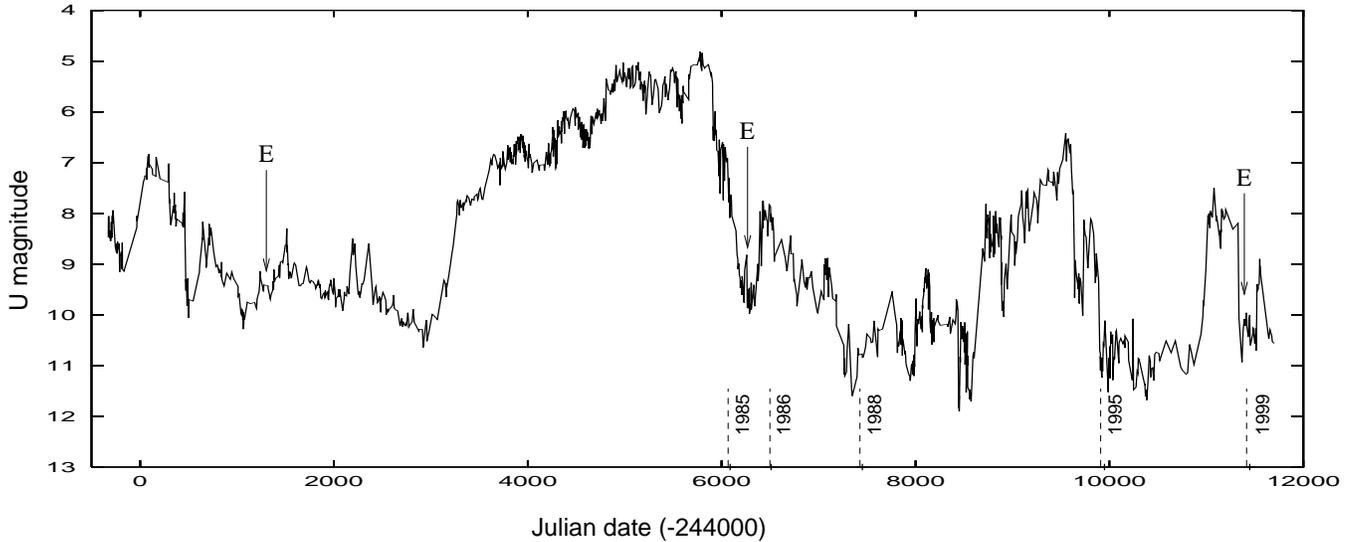}} \end{picture} \caption{The $U$ light-curve for
 CH\,Cyg covering the period 1967 -- 2000. The eclipses in the long,
 14.5-year, period orbit are marked by {\bf E}. The timings of our VLA
 observations pertinent to this paper are denoted by dashed bars.}
\label{fig:ulight}
\end{figure*}

\begin{table*}
\caption{VLA Observations of CH Cygni.}
\begin{minipage}{\linewidth}
\renewcommand{\thefootnote}{\thempfootnote}
\begin{tabular}{|l |l |l |l | l| l| l|}
\hline
Date        &Band 1&Band 2&Time on Source&Map 1noise\footnote{referred to as $\sigma_{\rm rms}$}&Time on Source&Map 2 noise\\  
            &(GHz) &(GHz) &at Band 1(s)  &(mJy beam$^{-1}$)                                     &at Band 2(s)  &(mJy beam$^{-1}$)\\
\hline    
1986 Mar 20 &4.8   &15    &2290          &0.053                                                 &9930          &0.083\\  
1988 Oct 16 &4.8   &15    &2600          &0.032                                                 &5220          &0.073\\   
1995 Aug 20 &4.8   &8.5   &2550          &0.021                                                 &5620          &0.063\\ 
1999 Sep 26 &4.8   &8.5   &3210          &0.077                                                 &2400          &0.133\\  
\hline
\end{tabular}
\end{minipage}
\label{tab:obs}
\end{table*}

The activity of CH\,Cyg is seen best in the $U$ light--curve (Skopal,
1998), which indicates a  brightening from $U \sim$11-12 to 5
(Fig. \ref{fig:ulight}). The drop in the optical magnitude of
CH\,Cygni  at the times of the long--period eclipses have been marked
on this figure.  Each outburst was accompanied by high velocity
broad emission features consistent with mass outflows. During 1984-85
the material was ejected  at $\sim$600-2500\,km\,s$^{-1}$
(Miko{\l}ajewski \& Tomov 1986), whilst the 1992-95 active phase was
characterized  by sporadic and in part bipolar outflow at
$\sim$1000-1600\,km\,s$^{-1}$ (Leedj\"{a}rv \& Miko{\l}ajewski, 1996;
Skopal et al. 1996; Ijima 1996) and, finally, during the recent,
1998-2000, outburst mass outflows at about 1000\,km\,s$^{-1}$ were
observed (Paper II). The outflows may be correlated with a
significant increase of the radio emission and the radio light curves
during these periods correlate well with the optical ones (Kenny et
al. 1996). The 1984 mass ejection has been linked  to emergence of
bipolar emission which could be from either an expanding torus or
collimated jets  (Taylor et al. 1986). Later radio observations
revealed a complex morphology, although still exhibiting a bipolar
structure (Kenny et al. 1996).

In this paper we present radio surface brightness and spectral index
maps taken from 15 years of observations of CH Cygni by the Very Large
Array.

\section{Observations}

CH Cygni has been observed in the highest resolution ``A''
configuration several times with the VLA since 1984 and multiple
frequency observations were made on many of these occasions, four of
which are detailed in Table~\ref{tab:obs}.

Data reduction was carried out according to standard procedures using
the AIPS software package (Greisen, 1999). The flux density of the
phase reference source was determined, and the data calibrated to
correct phase and amplitude instabilities.  These corrections were
applied to the observations of CH Cygni and a map produced by
deconvolving the instrumental response (the so-called ``dirty beam'')
from the map using the well-established CLEAN algorithm (H\"{o}gbom,
1974).  Due to the low radio flux density of the source no
self--calibration was possible. The maps were convolved with a beam
size of 0.5$\times$0.5~arcsec at all observing bands. This size was
the mean of the the major and minor axes of the beam at 5GHz, in order
that comparison of the size and shape of features at different epochs
would be meaningful. 
 
Observations of the same object at different frequencies have a
different $u,v$ plane coverage. This leads to features of different
spatial scales being picked out at the different frequencies, causing
problems when taking a ratio between two maps. In order to avoid this,
only the $u,v$ range covered by both sets of data was used to create
the maps. This has the undesirable effect of decreasing the
signal-to-noise ratio but does mean that the maps can be compared. The
AIPS task ``comb'' was  then used to produce a third map (at a
resolution of 0.17 arcseconds per pixel) which traces the
pixel--by--pixel spectral index variation. We use the convention that
the radio flux at frequency $\nu$, $S_\nu \propto \nu^\alpha$ so that
a positive spectral index ($\alpha$) represents an increase in flux
with frequency.

\begin{figure}
\begin{picture}(180,180)
 \put(0,0){\includegraphics{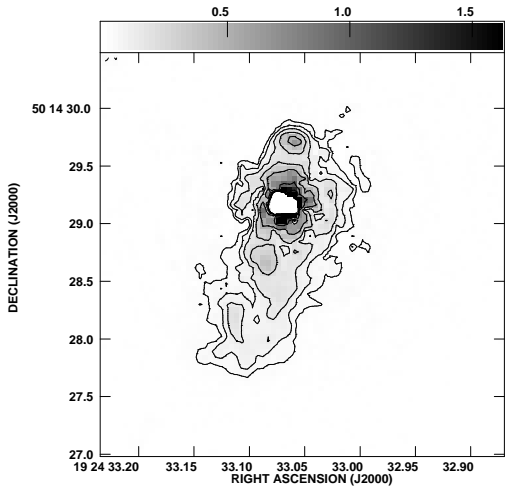}} \end{picture} \caption{HST WFPC2 image
 of CH Cygni from 1999 August 13. The point spread function of the
 central star has been subtracted from this image for clarity.}
\label{FIG:HST}
\end{figure}

Hubble Space Telescope imaging was used to place the stellar
components on the radio map. The observations were made as part of a
GO programme 8330 on symbiotic stars. Three orbits were allocated to
CH Cygni. A WFPC2 image (Fig. \ref{FIG:HST}) taken through the F502N
filter (Biretta 1996) on 1999 August 13 was aligned to  the VLA map of
1999~September. The best agreement was achieved when the radio  peak
coincided with the star in the optical domain (this is discussed in
more detail in Paper II).

\section{Results}

\begin{figure*}
\begin{picture}(600,600)
 \put(0,0){\includegraphics{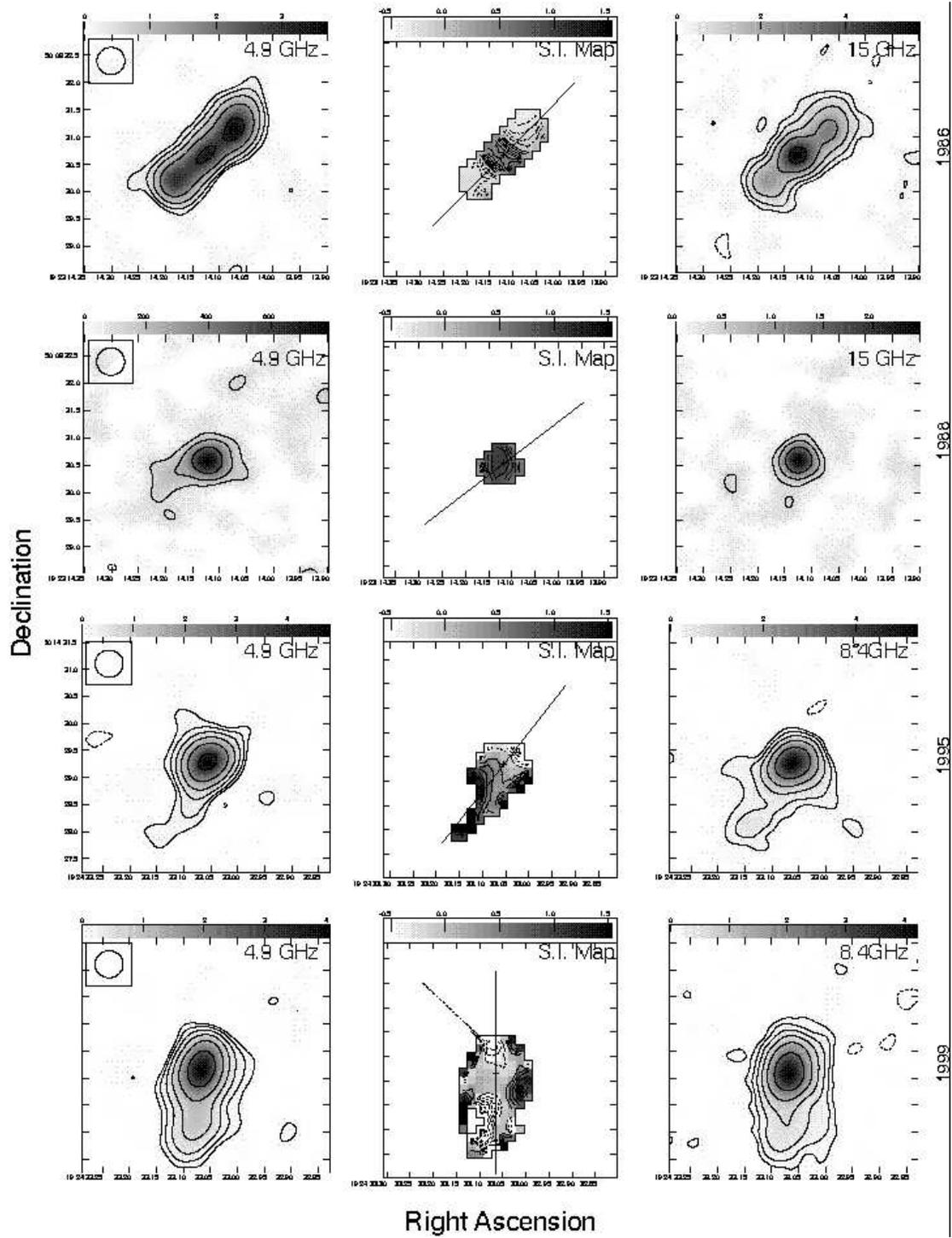}} \end{picture} \caption{VLA spectral index maps
 (centre) created from 4.8GHZ (left) and  8.5GHz or 15GHz (right) maps
 from from all epochs. The lowest contour line on the brightness maps
 is $\sigma_{\rm rms} \times -3, 3, 6, 12, 24, 48, 96$ where present. 
 The values of $\sigma_{\rm rms}$  for each map are given
 in Table ~\ref{tab:obs}. The white to black range on the spectral
 index map is  $-0.5 \leq \alpha \leq +1.5$ and the contours represent
 changes in $\alpha$ of 0.1.}  \label{FIG:maps}
\end{figure*}

The radio brightness and spectral index maps are shown in
Fig. \ref{FIG:maps}. The brightness maps are dominated by a central
bright peak surrounded by an extended region. The morphology of these
extensions varies from one epoch to the next. In particular, after the
initial bipolar extension observed in 1986 the extension appears to be
more pronounced on the SE side of the central peak. In addition to
this the position angle of the extension has changed (Crocker et al,
in preparation).

In each spectral index map, there is a wide range of indices
visible. In particular, the 1986 map (Fig. \ref{FIG:maps}, top) shows
a distinct trend with positive $\alpha$  in the central region and
negative $\alpha$ in the outlying areas. This strongly suggests the
existence of thermal emission around the stellar components with
extended non--thermal emission from the well-resolved knots.

\begin{figure}
\begin{picture}(200,600)
 \put(0,0){\includegraphics{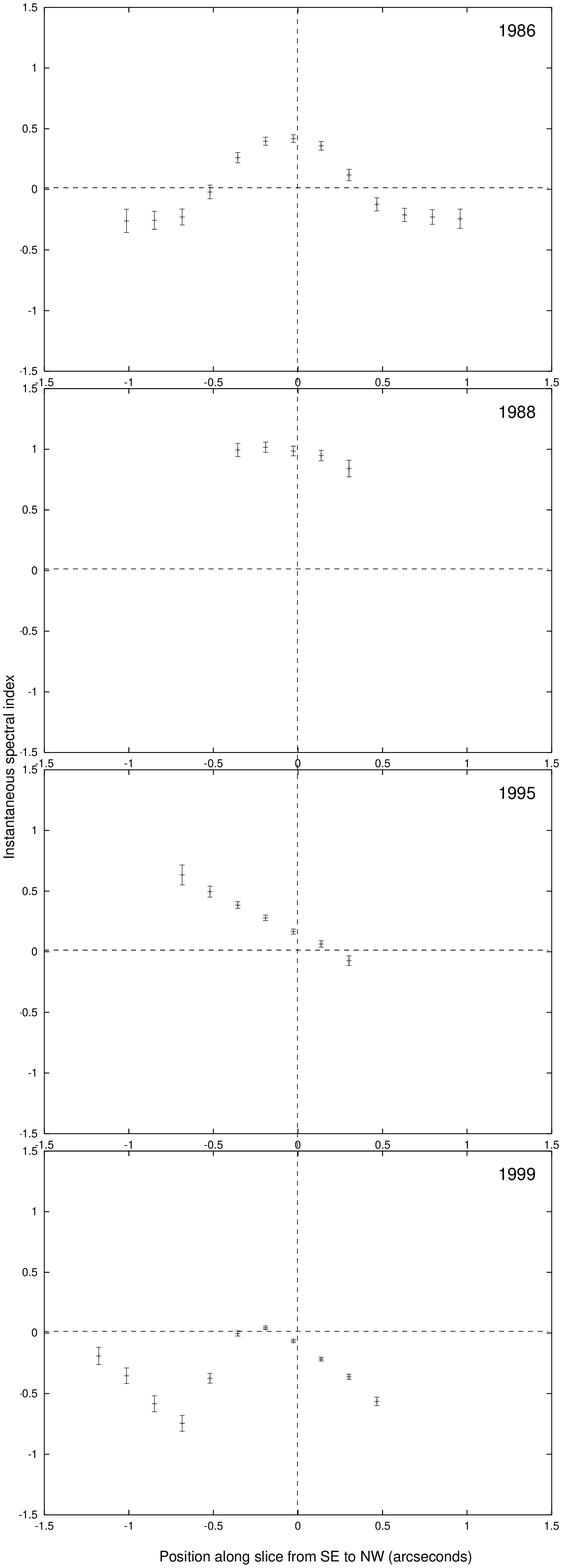}} \end{picture} \caption{Plots of instantaneous
 spectral index variations along the lines marked out in
 Fig. \ref{FIG:maps}, from SE to NW, for each of the maps, relative to
 the bright peak in the higher frequency radio brightness map.}
 \label{FIG:PLOT}
\end{figure}

Cross--sections were taken through the peak of each spectral index
map, along the axis of maximum elongation in the 5GHz map. In order to
reduce the uncertainties, a section was taken on each side of the
initial one, giving three parallel sets of data. At each point along
the slice, the three pixels were combined to find a weighted mean and
then used to create the final cross-section.  The weighted mean
cross-sections are shown in Fig. \ref{FIG:PLOT}, where the  positions
are given relative to the central bright peak in the 8.5GHz or 15GHz
radio maps, whichever was available. A feature common to all the
epochs is a central peak, accompanied  by wings of lower spectral
index.

The first plot, from 1986, has a peak value of $\alpha = 0.42 \pm
0.03$ which is consistent with that expected from a constant outflow
ionized wind, assuming a  variation in density of $\rho \propto
r^{-\beta}$, with $\beta = 1.8$. The decrease in $\alpha$ with
distance from the centre favours a geometry which is optically thick
along lines of sight that pass through a dense core. In the extended
regions, the mean value of $\alpha$ is $-0.23 \pm 0.03$.

By 1988, the spectral index of the central region had increased to
$\alpha = 1.02 \pm 0.04$ as if $\beta>2$, which can be attributed to
the photoionised nebula dominating over the wind. Such a value is
predicted by the STB model for lower ionisation parameters. No extended 
emission is seen at 15GHz. This observation was made at a time when the 
optical activity was very low (Fig. \ref{fig:ulight}) and no outburst 
had been observed for some time.

The 1995 cross-section has a peak of similar width to the other epochs
in conjunction with  bright thermal emission ($\alpha > 0.5$) in the
extension seen to the SE of the  central source in the 8.5GHz map in
Fig. \ref{FIG:maps}. This feature may be a result of local
condensation in the mass flow.

In 1999, almost the entire emitting region has a negative spectral
index, with a minimum of $\alpha = -0.75 \pm 0.07$. This observation
was made during what may be an eclipse of the inner symbiotic region
by the outer giant (Skopal 1997). This may have reduced the relative
intensity of the thermal emission,  suggesting that the thermal
emission region was very compact at this time, i.e. similar in extent
to the angular diameter of the outer red giant.

\section{Discussion}

\subsection{The geometry of the extended emission}

The extensions detected in the radio surface brightness maps are more
likely to be collimated jets rather than a torus. A torus is unlikely
to appear so obviously one--sided, following an initial bipolar
morphology. The non--thermal emission is suggestive of high velocity
flow, commonly associated with jets but not with a circumbinary torus.

It may be that the extended emission is due to a large nebula, with 
a radius of order 100AU, which has been photoionised between 1985 and 1986.  
From the STB model, for
suitable parameters ($T_{\rm H}=100,000$K, $\dot{M}= 1.9\times
10^{-6}~M_\odot$yr$^{-1}$ (Skopal et al., 1996), $V_{\rm
wind}$=20~km~s$^{-1}$ (Vogel et al., 1994)) it is clear that the
ionised cavity due to the hot component will be density bounded and
much smaller than the binary separation.

\subsection{Expansion velocity}

The Hipparcos distance to CH Cygni is $268\pm66$~pc (Viotti et
al. 1998), meaning that the maximum  extent of the $3\sigma_{\rm rms}$
contour of radio emission at 15GHz in 1986 is  $\sim 500$AU. The most
likely origin of of extended non--thermal emission is synchrotron
emission from relativistic electrons in a magnetic field.

If CH Cygni follows the model proposed for HM Sagittae by Eyres et
al. (1995), then  what is being observed are regions of shocked
material, caused when high velocity ejecta expand into the relatively
dense circumtrinary envelope created by the winds of the giant stars
in the system.

Evidence for this scenario comes from infrared spectroscopy (Taranova
\& Yudin, 1988), carried out around the period of the end of the
optical outburst in 1983--1987, showing changes in the circumtrinary
dust envelope. They suggested that the dust was disrupted at the time
of the radio jet emergence, probably as it was swept up by the
expanding matter.

\begin{figure}
\begin{picture}(180,180)
 \put(0,0){\includegraphics{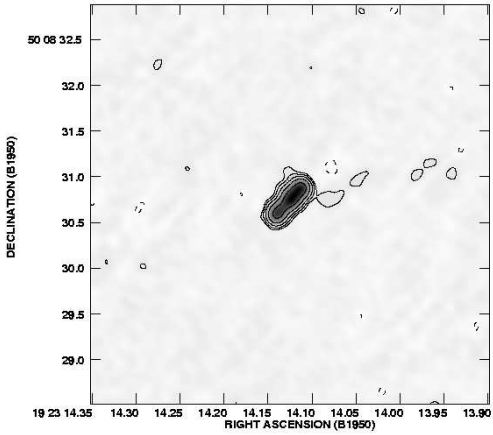}} \end{picture} \caption{VLA 15GHz map from 1985
 January 22. The contours are as in Fig \ref{FIG:maps} and the rms
 noise level is 0.19mJy per beam.}  \label{fig:85}
\end{figure}

The positions of the NW and SE extended knots were measured on the
1986 15GHz radio  map. The angular separations between the peak in
each knot and the central bright peak were

$$
\Delta\theta_{\rm NW} = 0.70 \pm 0.01 {\rm ~arcsec}
$$

\noindent and

$$
\Delta\theta_{\rm SE} = 0.71 \pm 0.05 {\rm ~arcsec}
$$

\noindent for the NW and SE knots respectively. The greater
uncertainty in the SE position is due to the fact that the SE
component has lower surface brightness than the the NW one. In a 1985
VLA 15GHz map (Fig. \ref{fig:85}), these two knots were not visible,
and hence the material must have been ejected within 422~days before
the 1986 observation. This places a lower limit on the  expansion
velocity. Assuming a distance to CH Cyg of $268\pm 66$~pc the
velocities  are found to be

$$
V_{\rm NW} \geq 1210 \pm 300 {\rm ~km~s}^{-1}
$$

\noindent for the NW component and

$$
V_{\rm SE} \geq 1230 \pm 320 {\rm ~km~s}^{-1}
$$

\noindent for the SE component, with the uncertainties dominated by
the  uncertainty on the Hipparcos distance. These results correspond
well with those derived previously, in particular with the expansion
velocity of 1400~km~s$^{-1}$ derived from two earlier maps (Taylor et
al. 1986) that were separated by only 75 days.

\subsection{Magnetic field in 1986}

The upper limit of the surface  magnetic field of a typical white
dwarf is around $10$G (Bond and Chanmugam, 1982). Assuming a dipole
geometry, this  will fall off as $B_{\rm WD} \propto r^{-3}$. For a
stellar radius of 0.01R$_\odot$ the upper limit on the magnetic field
at 500AU is $80~\mu$G, which is far too low to give efficient
synchrotron emission.

The speed of the unshocked pre--existing cool wind is likely to be
much less than the jet speed, as a typical cool giant wind has a
velocity of the order of 20--40\,km~s$^{-1}$ (Vogel et al. 1994). The
non--thermal emission is then caused by particle acceleration at the
resulting bow shock, as the local magnetic field is compressed and
enhanced and electrons are accelerated to velocities close to $c$. In
such an environment, the local azimuthal field, $H_\phi$ is defined by
Equation 42 of Bode \& Kahn (1985) the density of the shock, the 
temperature of the giant wind (assumed to be constant at $\sim 10^4$K) 
and the mean particle mass which is taken to be the proton mass. 
The shock is assumed to be strong and totally
ionised, so that the shock density is equal to 4 times the density
of the unshocked wind at this distance from the star. It is assumed
that the disruptions caused to the giant wind by the hot component
prior to mass ejection are negligible at this distance so that a
uniform, spherical radial wind exists.

The denisty of the wind is proportional to the mass loss rate of the
giant, $\dot{M} = 1.9\times 10^{-6} {\rm M}_\odot~{\rm yr}^{-1}$  (Skopal,
1996). If the unshocked wind speed, $v_w$ is taken to be the escape
velocity of the red giant, assuming $M_{\rm giant} = 0.93M_\odot$ and
$R_{\rm giant} = 123R_\odot$ (Skopal 1997) then $v_w  = 53.8~{\rm
km~s^{-1}}$ and $H_\phi\simeq 2.4$~mG.

Another magnetic field estimate can be derived from the technique used
in studying high energy extragalactic sources. The total energy of a
radio source is split between that of the electrons, the protons and
the magnetic field. At equipartition, the magnetic energy is
approximately equal to the particle energy (Pachoczyk, 1970), and the
ratio of energy carried by protons to that carried by electrons is
$K$. For a radio source of known angular dimensions and depth, assuming an 
ellipsoidal geometry, the magnetic field under equipartition, $H_E$, can 
be found using the method described by Miley (1980).

To make use of this model, one needs to determine several parameters
that describe the region: filling factor, $\eta$; the angle between
the magnetic field and the line of sight, $\chi$; the flux density of
the emitting region, $S_\nu(Jy)$, at frequency $\nu$ (GHz) and the
maximum and minimum frequencies of radio emission ($\nu_{\rm max}$ and
$\nu_{\rm min}$ repectively, in GHz) from the source spectrum.

The filling factor would be more useful in extragalactic work when it
would be expected that there would be unresolved blobs of emitting
material. Since the measured sizes of the knots in the extended emission
are of a similar angular size to the beam, $\eta = 1$. The angle $\chi$
is important due to the fact that the radiation observed is beamed
from electrons moving towards the observer. The fact that eclipses are
seen in the system means that the  orbital plane of the stars is
observed edge--on. Bipolar jets are believed to be aligned with the
rotation axes of their parent star (perpendicular to the orbital
plane) so the jets of CH Cygni  are believed to lie almost in the
plane of the sky (Crocker et al., in preparation),  $\phi \simeq
90^\circ$. The cutoff frequencies of the radio emission, $\nu_{\rm
min}$ and $\nu_{\rm max}$ are taken to be 0.01GHz and 100GHz
respectively (Miley, 1980). The value of $K$ is the main unknown in
this equation as the exact details of the mechanism leading to
synchrotron emission are not known. A value of $K=100$ would be
appropriate for the electrons being produced following collisions in a
circumstellar medium (Pacholczyk 1970), but a value of $K=1$ is more
consistent with the minimum energy requirement (Miley 1980).  However,
since $H$ only depends upon $K^\frac{2}{7}$, having $K$ 100 times
larger would only increase $H_E$ by a factor of 3.

In order to determine the size of the jet bullets, two dimensional
gaussians were fitted to the emitting regions in the NW and SE
jets. The depth was then found by assuming that the regions exhibited
cylindrical symmetry along the axis of ejection. For the NW jet,
$\theta_1 = 0.39$ arcsec, $\theta_2 = 0.34$ arcsec and $d = 91$
AU. The SE jet had dimensions of $\theta_1 = 0.37$ arcsec, $\theta_2 =
0.35$ arcsec and  $d = 94$ AU. Using the mean value of the spectral
index in the jets of $\alpha = -0.23 \pm 0.03$,

$$H_E = 3.8~{\rm mG}$$

\noindent for the NW jet and

$$H_E = 3.6~{\rm mG}$$

\noindent in the SE. These values are comparable to that found above
from the physics of the shock, so there is sufficient magnetic field
in the enhanced regions for synchrotron radiation.

\subsection{Lifetimes}

Since the total luminosity of this region at all frequencies is not
known, determining the lifetime of the emission due to all radiative
energy losses is impossible. In addition to this, synchrotron loss
processes may be dominant. Under the influence of a magnetic field
$H$, an electron rotates at the gyrofrequency (Rybicki \& Lightman,
1979) and the frequency of radiation, $\nu$, is a factor of $\gamma^3$
higher than the  gyrofrequency, so manipulating Equation 6.4 of 
Rybicki \& Lightman (1979) gives $\gamma =1148$. An electron  with this 
Lorentz factor has energy $\gamma m_e c^2$ and radiates with power $P$ 
so that the lifetime of the synchrotron radiation is
$$
\tau = \frac{9m_e^3c^5}{4\gamma e^4H^2}
$$
\noindent and with the values of $\gamma$ and $H = H_\phi$ from above, $\tau =
4.7\times 10^{10}$ seconds,  corresponding to just under 1500
years. Such a timescale would imply that synchrotron radiation  would
be very long-lived in the circumstellar material, assuming that the
magnetic field compression is sustained by continuous matter
injection. When there is no matter injection the magnetic field can
expand adiabatically and the synchrotron emission will fade within
years. No non-thermal emission is seen in 1988.  If indeed there is no
synchrotron emission at this epoch, some other loss-mechanism must be
occurring. The optical light curve of figure \ref{fig:ulight} shows
1988 to be a particularly quiet period, over two years after the last
major outburst. It is therefore likely that there is no matter being
ejected into the circumstellar wind, so that no new density
enhancements and magnetic field  compressions are being formed.

Supporting evidence for this scenario comes from optical spectral
observations. During this  epoch, the blue continuum was very faint
and the Balmer lines were not present (Bode et al. 1991),  strongly
implying that there were also no fast winds or jets at this
time. Hence one would  not expect to detect shock phenomena resulting
from penetration of high--velocity material into  circumbinary
material and, by this model, no non--thermal emission from recently
ejected material.

\subsection{Comparison with other symbiotic stars}

Previously, unambiguous non--thermal emission in symbiotic stars had
only been seen in HM Sge (Richards et al. 1999).  Spatially resolved
non--thermal emission is very rare in any stellar source, and is only
seen regularly in very high--energy systems such as micro-quasars. A
potential non--thermal feature in R Aqr was subject to such large
errors ($\sigma_\alpha \sim 1$) that its status could not be confirmed
(Dougherty et al. 1995). Other symbiotic stars are believed  to
exhibit entirely thermal radiation. CH Cyg displays the strongest
evidence yet of non--thermal emission due to jet--wind interaction.
CH Cyg and HM Sge share the property of having undergone outbursts in
relatively recent history. The above arguments suggest that
non--thermal emission will decrease rapidly with time after the
outburst if the field strength and high energy particles are not
continuously replenished as discussed initially in Richards et
al. (1999).  Since the 1984-86 ejections in CH Cygni two outbursts
have been observed  which may have injected further high velocity
material into the circumstellar environment and hence given rise to
the non-thermal emission seen in the 1995 and 1999 observations.

\section{Conclusion}

There is definite evidence of non--thermal emission in the ejecta from
CH Cygni. This, together with a morphology consistent with high
velocity jets, supports the  scenario of Taylor et al. (1986), rather
than one where the extension is a result of an expanding circumstellar
torus. The proposed mechanism for the creation of the non-thermal
emission involves  material ejected in high velocity bullets
interacting with a previously expelled lower velocity stellar
wind. The ejected matter creates shocks in the circumstellar medium,
leading to locally enhanced magnetic field density, particle
acceleration and synchrotron emission. We suggest that the presence 
of non--thermal emission in symbiotic stars may be an indicator of 
outbursts associated with significant ejecta.

\section*{Acknowledgments}

MC is supported by a grant from the Particle Physics and Astronomy
Research Council (PPARC).  The contribution of AS was supported by a
grant of the Slovak Academy of Science, number 5038/2000. The VLA is
operated by the National Science Foundation operated under cooperative
agreement with Associated Universities, Inc.

\label{lastpage}
\end{document}